\documentclass{article}
\usepackage[utf8]{inputenc}
\usepackage{geometry}
\usepackage{graphicx}
\usepackage{hyperref}
\usepackage{multirow}
\usepackage{tabularx}
\usepackage{color}
\usepackage{amsmath}
\usepackage{amssymb}
\usepackage{amsfonts}
\usepackage{amsxtra}
\usepackage{wasysym}
\usepackage{isomath}
\usepackage{mathtools}
\usepackage{txfonts}
\usepackage{upgreek}
\usepackage{enumerate}
\usepackage{tensor}
\usepackage{pifont}
\usepackage{footmisc}
\usepackage{color}
\usepackage{verbatim}
\usepackage{amsmath}
\usepackage{indentfirst}
\usepackage{mathrsfs}
\usepackage{algorithm2e}
\usepackage{dcolumn}
\usepackage{bm}

\usepackage{color}
\usepackage{comment}
\title{NP-Hardness of Tensor Network Contraction Ordering}

\author{Jianyu Xu\footnote{for equal contribution}, Hanwen Zhang$^{*}$, Ling Liang, Lei Deng, Yuan Xie, Guoqi Li
}

\begin{document}
\bibliographystyle{unsrt}
\maketitle

\begin{abstract}
	We study the optimal order (or sequence) of contracting a tensor network with a minimal computational cost. We conclude 2 different versions of this optimal sequence: That minimize the operation number (OMS) and that minimize the time complexity (CMS). Existing results only shows that OMS is NP-hard, but no conclusion on CMS problem. In this work, we firstly reduce CMS to CMS-0, which is a sub-problem of CMS with no free indices. Then we prove that CMS is easier than OMS, both in general and in tree cases. Last but not least, we prove that CMS is still NP-hard. Based on our results, we have built up relationships of hardness of different tensor network contraction problems.
\end{abstract}

\section{Introduction}

Tensor network \cite{orus2014practical} is important in quantum mechanics \cite{biamonte2017tensor} \cite{biamonte2013tensor}, high-dimentional data analysis \cite{cichocki2014tensor} \cite{xu2018towards} and artificial intelligence \cite{socher2013reasoning}. It also has a large number of applications, including quantum computing \cite{markov2008simulating} \cite{boixo2017simulation} \cite{chen2018classical} \cite{arad2010quantum}, machine learning \cite{novikov2015tensorizing} \cite{yang2017tensor}, signal processing \cite{cichocki2015tensor} and quantifications \cite{biamonte2015tensor}. However, to computing a tensor network asks for large amount of computational cost. Traditionally, we compute a tensor network by doing contractions on a pair of tensors for each time, until there exists only one tensor. Though contraction sequences do not affect the computational result, different sequences of carrying out contractions may lead to huge variance of computational cost, which is similar to matrix chain products. Therefore, it is important to determine the optimal sequence of contractions.

However, this is not a easy task at all. On the one hand, the work in \cite{chi1997optimizing} proved that it is NP-hard to determine a sequence of contractions that has minimal computational cost (the number of multiplications of elements among the whole contraction procedure). On the other hand, the work in \cite{xu2019towards} proposed a polynomial algorithm that determines the optimal time complexity (the largest number of multiplications of elements among every single pairwise contraction, which is also adopt in \cite{fried2018qtorch}) on any tensor tree algorithm. Other works such as \cite{pfeifer2014faster} and \cite{liang2020fast} contributed to prompting the efficiency on finding the optimal sequence, but they did not determine the intrinsic hardness of this problem. 

Therefore, a huge gap occurs between \cite{chi1997optimizing} and \cite{xu2019towards}. Primarily, they dealt with different problem settings, the former finding minimal computational cost while the latter determining optimal time complexity. Thus, a vital question occurs: is the problem on computational cost harder than that on time complexity? Additionally, the \cite{xu2019towards} work only dealt with tensor tree networks. Is the problem of determining a sequence of optimal time complexity in general cases much harder than that in tree cases? In other words, is the problem still polynomial on general tensor networks, or is it an NP-hard one?

In this work, we have completely answered the two questions above. Firstly, we point out that the problem setting on computational cost is much harder than that on time complexity, by proving that the former problem is still NP-hard even on tensor tree networks while the latter has already proved to be polynomial in \cite{xu2019towards}. Secondly, we prove that the determination of sequence of contractions for optimal time complexity is an NP-hard problem on general tensor networks. With these two problem being answered, we have build up the framework of hardness of optimal sequence problems on tensor network contractions. Also, for further research on determining the optimal sequence of contracting a specific tensor network, it is easier to consider time complexity instead of computational cost.

\section{Definitions and Terminologies}
\textbf{Tensor Contraction}\cite{xu2019towards}: For a number of tensors, we name the operation \textit{tensor contraction} when we sum over some common indices, each pair of which occurs and only occurs twice, of these tensors as inner products while remaining the other indices as outer products, each of which occurs and only occurs once. We name these common indices \textit{dummy indices}, and the other indices \textit{free indices}.

For example: the following calculation:
$$\left(AB\right)_{a,d,e}=\sum_{b=1}^{N_b}\sum_{c=1}^{N_c}A_{a,b,c,d}\cdot B_{b,c,e}$$
is a contraction of tensor $A$ and $B$, where $b$ and $c$ are dummy indices and $a, d, e$ are free indices.

In the following parts, without specifically mentioned, ``tensor contraction'' refers to that of 2 tensors.

\textbf{Tensor Network}\cite{xu2019towards}: For a set of many tensors, there are relationship of multiplications between pairs of tensors. In order to illustrate these relationship, we use a graph to represent their relationship: we use vertices to represent tensors, and for any pair of tensors, we connect them with an edge if they share some common indices. \textbf{Weights} are given to vertices and edges in order to show the ``order'' (or ``mode'') of tensors and multiplications between pairs of tensors. We call this graph a \textit{tensor network}.

%According to the definition given by , ``In a contraction of many tensors, we use a graph to represent their relationship: For every tensor, we use a vertex $A$ to represent it, and give it a \textit{weight} $W_{A}$ that equals the number of free indices it has. For every pair of tensors (vertices) $A$ and $B$, we use an edge $E_{A-B}$ to connect them, and we give it a weight $W_{A-B}$ that equals the number of dummy indices they have in common. For every edge whose weight is 0, we {\rm{can}} delete it. We call this network a \textit{tensor network}, usually denoted as $TN$ (or $T$ for a tree structure in the following sections).''

\textbf{Tensor Network Contraction}: According to the definition given by \cite{xu2019towards}, we can transform the definition of tensor contraction to a tensor network: (1) We draw another vertex to represent the result of contraction of the selected tensors (vertices), and contract their weights as that of the new vertex. (2) For every edge who has one and only one end in the selected vertices, we move this end to the newly drawn vertex, with its weight unchanged. (3) For any pair of vertices between whom there is more than one edge, we ``merge'' these edges into one edge and contract their weights as that of the new edge. (4) Erase those selected vertices and edges between them.''

\begin{figure}
    \centering
    \includegraphics[width=0.9\textwidth] {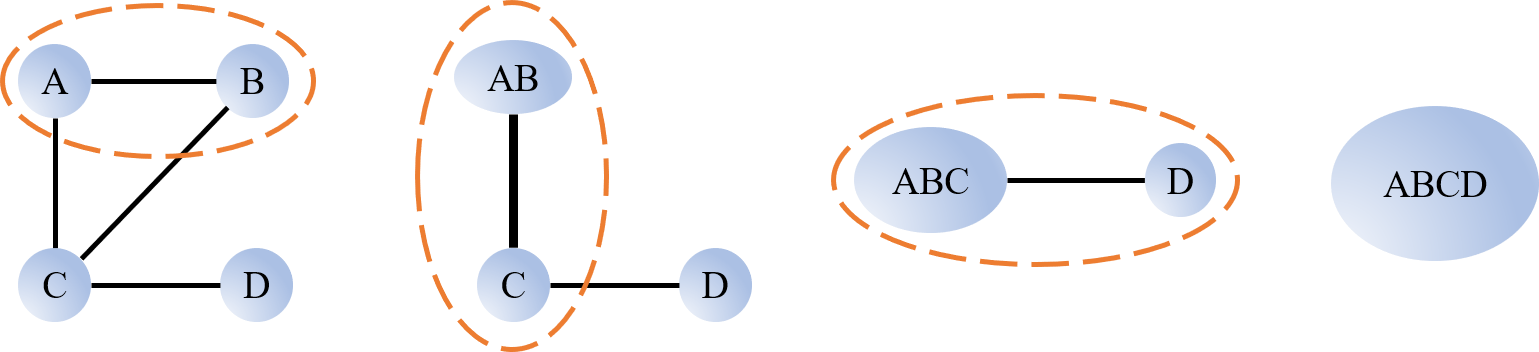}
    \caption{Representations of Equation \ref{TNC} in tensor network form and its contractions.}
    \label{Tensor Network}
\end{figure}

For example, we can use tensor network contraction in Figure \ref{Tensor Network} to represent the following equations:

\begin{equation}
\label{TNC}
\begin{aligned}
{ABCD}_{i,j,k}&= \sum_{l=1}^{N_l}\sum_{p=1}^{N_p}\sum_{q=1}^{N_q}\sum_{r=1}^{N_r}\sum_{u=1}^{N_u}\sum_{v=1}^{N_v}\sum_{w=1}^{N_w}A_{ilpq}\cdot B_{lpr}\cdot C_{jqruvw}\cdot D_{kuvw}\\
&= \sum_{q=1}^{N_q}\sum_{r=1}^{N_r}\sum_{u=1}^{N_u}\sum_{v=1}^{N_v}\sum_{w=1}^{N_w} C_{jqruvw}\cdot D_{kuvw}(\sum_{l=1}^{N_l}\sum_{p=1}^{N_p}A_{ilpq}\cdot B_{lpr})\\
&= \sum_{q=1}^{N_q}\sum_{r=1}^{N_r}\sum_{u=1}^{N_u}\sum_{v=1}^{N_v}\sum_{w=1}^{N_w} C_{jqruvw}\cdot D_{kuvw} \cdot(AB)_{iqr}\\
&= \sum_{u=1}^{N_u}\sum_{v=1}^{N_v}\sum_{w=1}^{N_w} D_{kuvw} (\sum_{q=1}^{N_q}\sum_{r=1}^{N_r}C_{jqruvw}\cdot(AB)_{iqr})\\
&= \sum_{u=1}^{N_u}\sum_{v=1}^{N_v}\sum_{w=1}^{N_w} D_{kuvw}\cdot(ABC)_{ijuvw}\\
&= (ABCD)_{ijk}
\end{aligned}
\end{equation}

From Equation \ref{Tensor Network}, if we would like to achieve the computational result, we need to conduct all summations of the indices. Since these summations are independent to each other, different sequences of conducting these summations would lead to the same result. However, there could be a lot of difference on their computational cost. For example, consider the following matrix-vector product:
$$ABx,\ where\ A, B\in \mathbb{R}^{n\times n},\ x\in \mathbb{R}^n$$
. If we firstly compute $AB$ and then multiply it with $x$ on the left, then the time complexity will be $O(n^3)$. But if we first compute $Bx$ and then multiply $A$ with the vector, then the time complexity will be $O(n^2)$, which is much less than the former one. Therefore, it is important to reduce and even minimize the computational cost for tensor network contractions, by determining an optimal sequence of contractions. Now we are faced with 2 problem: (1) how to represent the difference, and (2) how to find out the least cost, as well as a corresponding sequence. To make this, we propose the following definitions:

\textbf{Multiplicational Representation} of a tensor network: For any vertex or edge, its weight is given as the \textbf{product of all dimensions involved} in the corresponding place. In this representation, contractions of weights means to multiple them together.

\begin{figure}
    \centering
    \includegraphics[width = 0.9\textwidth]{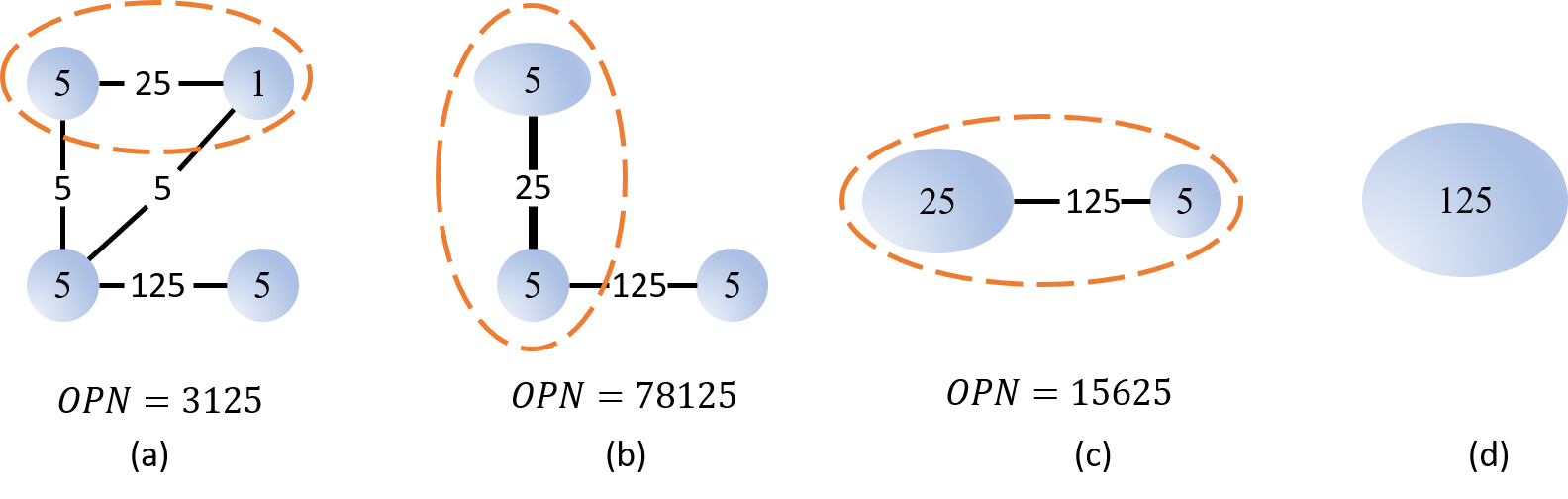}
    \caption{A multiplicational representation of Equation \ref{TNC}. Here we suppose $N_i = N_j = \ldots = N_v = N_w = 5$.  For each step, the operation number equals a product of all weights involved. The total operation number equals the summation of that of each steps, which is $3125+78125+15625 = 96875$.}
    \label{MR}
\end{figure}

%For example, suppose $A \in \mathbb{R}^{m\times n\times k}, B\in \mathbb{R}^{n\times l}$, then $A$ multiplied with $B$ can be represented by a tensor network: vertices $V_A, V_B$ and edge $e_{AB}$. In a multiplicational representation, their weights are $W_A=m\times k$, $W_B=l$ and $W_{A-B}=n$ sequentially, and the result $V_{AB}$ of this contraction has a weight of $m \times k \times l$.\\ 
An example comes with Equation \ref{TNC} and its corresponding multiplicational representation in Figure \ref{MR}. Here we suppose $N_i = N_j = \ldots = N_v = N_w = 5$.  For each step the operation number equals a product of all weights involved, and the total operation number equals the summation of that of each steps.

\textbf{Additional Representation} of a tensor network: For any vertex or edge, its weight is given as \textbf{the number of orders involved} in the corresponding place. Equivalently, any weight in additional representation equals the \textbf{logarithm} of that in multiplicational representation. In this representation, contractions of weights means to add them together.

\begin{figure}
    \centering
    \includegraphics[width = 0.9\textwidth]{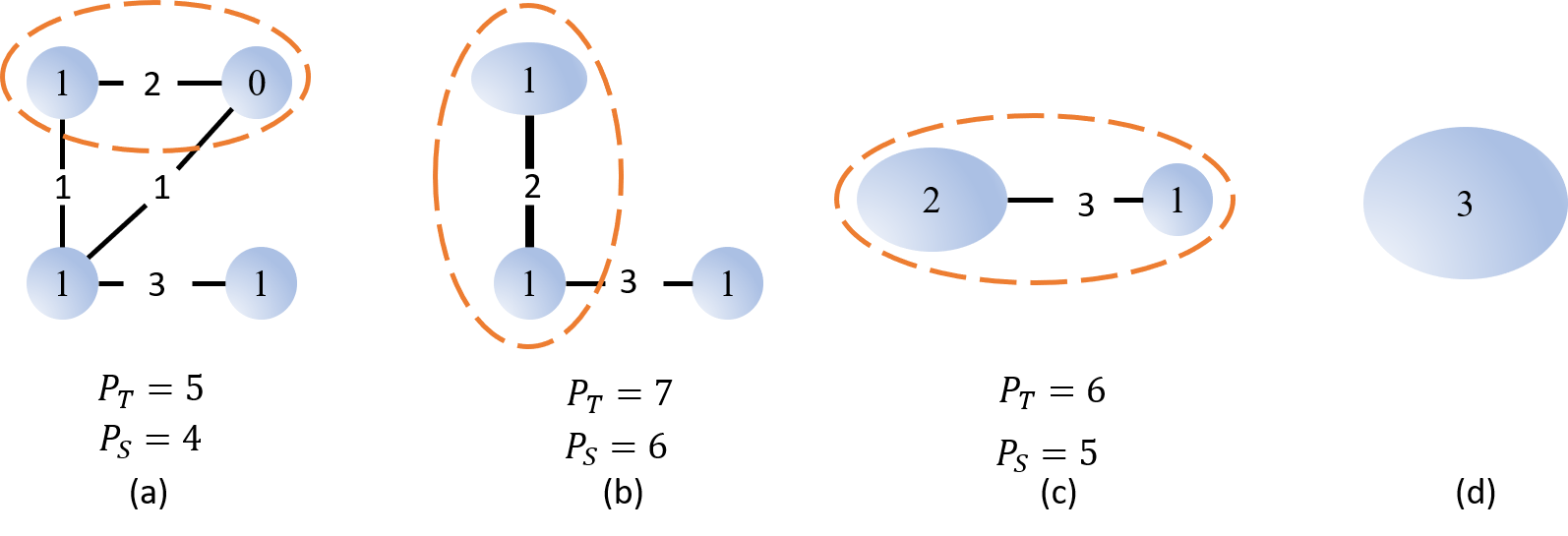}
    \caption{An additional representation of Equation \ref{TNC}. Suppose $N_i = N_j = \ldots = N_v = N_w$. For each step, the time power equals a summation of all weights involved, and the space power equals the maximum of $WD$ values of vertices being contracted and born from this step of contraction. For the whole sequence of contractions, the time power and space power equal to the maximum among all steps, correspondingly. Therefore, ${P_T}_{total} = \max\{5, 7, 6\} = 7$ and ${P_S}_{total} = \max\{4,6,5\} = 6$. Note that a maximum of $P_T$ does not always come together with a maximum of $P_S$.}
    \label{AR}
\end{figure}

%In the same instance proposed in the multiplicational representation, if we adopt additional representation then the weights would turn out to be $W_A=2, W_B=1, W_{A-B}=1$ and $W_{AB}=3$, given that $m=n=k=l$. If those dimensions are not equal, then we can adopt a logarithm with any base $N \geq 1$, then: $W_A=log_N(mk), W_B=log_N(l), W_{A-B}=log_N(n)$ and $W_{AB}=log_N(mkl)$.
We use the same example of Equation \ref{TNC} in comparison of its multiplicational representation. With the same assumption $N_i = N_j = \ldots = N_v = N_w = 5$, we can construct a tensor network and then conduct its contractions in Figure \ref{AR}. For each step, the time power equals a summation of all weights involved, and the space power equals the maximum of $\{WD(X), WD(Y), WD(XY)\}$, where $X$ and $Y$ are original vertices being contracted in this step, and $XY$ is the contraction result of $X$ and $Y$. For the whole sequence of contractions, the $P_T$ (or $P_S$) value equals maximum of $P_T$ (or $P_S$) for each step.

In this work, without specific mention, a ``Tensor Network'' refers to the additional representation of this tensor network.
%Degree, Weight and Degree to be defined

\textbf{Operation Number}: number of multiplications of scalars, denoted as $OPN$. For a single step of 2-tensor contraction, it equals the product of all weights, in multiplicational representations, of all vertices and edges involved in the contraction. For the whole process of contracting a tensor network, the number operations equals to the summation of operation numbers of all single steps.

\textbf{Time Complexity} and \textbf{Space Complexity}:

{For tensors $A\in \mathbb{R}^{M_1\times M_2\times\ldots\times M_m\times N_1\times N_2\times\ldots\times N_n}$ and $B\in \mathbb{R}^{ N_1\times N_2\times\ldots\times N_n\times Q_1\times Q_2\times\ldots\times Q_q}$, where $M_r, N_s, Q_t\in Z^{+}, \forall r=1,\ldots,m;s=1,\ldots,n;t=1,\ldots,q$, the time complexity of contracting $A$ with $B$ as the expression
\begin{equation}
\begin{aligned}
&(AB)_{i_1,i_2,\ldots,i_m,k_1,k_2,\ldots,k_q}\\
=&\sum_{j_1=1}^{N_1}\sum_{j_2=1}^{N_2}\ldots\sum_{j_n=1}^{N_n} A_{i_1,i_2,\ldots,i_m,j_1,j_2,\ldots,j_n}\cdot B_{j_1,j_2,\ldots,j_n,k_1,k_2,\ldots,k_q}
\end{aligned}
\end{equation}

equals $M_1 M_2 \ldots  M_m N_1 N_2 \ldots  N_n  Q_1  Q_2 \ldots Q_q$, and the space complexity of that equals
\begin{equation}
\begin{aligned}
\max&\left\{M_1 M_2 \ldots  M_m N_1 N_2 \ldots  N_n ,\right.\\
&\left. N_1  N_2 \ldots  N_n Q_1  Q_2 \ldots Q_q,\right.\\
&\left.M_1 M_2 \ldots M_m Q_1 Q_2 \ldots Q_q\right\}.
\end{aligned}
\end{equation}
Specifically, if $M_r=N_s=Q_t=N$, the time complexity equals $N^{m+n+q}$ and the space complexity equals $N^{\max\left\{m+n,n+q,m+q\right\}}$.
}

\textbf{Note}: For a sequence of contracting a tensor network, the time/space complexity equals the maximum time/space complexity among every single step.

\textbf{Time Power ($P_T$)}:

In a contraction of two vertices, we use $P_T$ to express time power:
\begin{equation}
P_T(AB)=WD(A)+WD(B)-W_{A-B}\label{timepower},
\end{equation}
where $W_{A-B}$ means the weight of edge connecting $A$ and $B$, and $WD(A)\overset{\Delta}{=} W(A)+\sum_{B\in V(G)}W_{A-B}$ denotes the sum of \textbf{Weight and Degree}.

\textbf{Space Power ($P_S$)}:

In a contraction of vertices, we use $P_S$ to express the space power.
\begin{equation}
P_S(AB)=\max\left\{WD\left(A\right), WD\left(B\right),WD\left(AB\right)\right\}\label{spacepower}.
\end{equation}

\textbf{Group of Vertices}: for several vertices that are contracted into the same vertex, we call this vertex a Group of those contained vertices, and any of those vertices a \textbf{Member} of the Group.

Based on the different scales of computational cost, we can define the following three problems:

\textbf{OMS}: the problem of determining operation-number-minimum contraction sequence of a tensor network \cite{chi1997optimizing}.

\textbf{CMS}: the problem of determining time-complexity-minimum contraction sequence of a tensor network \cite{xu2019towards}.

\textbf{CMS-0}: the problem of determining time-complexity-minimum contraction sequence of a tensor network with all vertices weight zero.

\begin{figure}
    \centering
    \includegraphics[width = \textwidth]{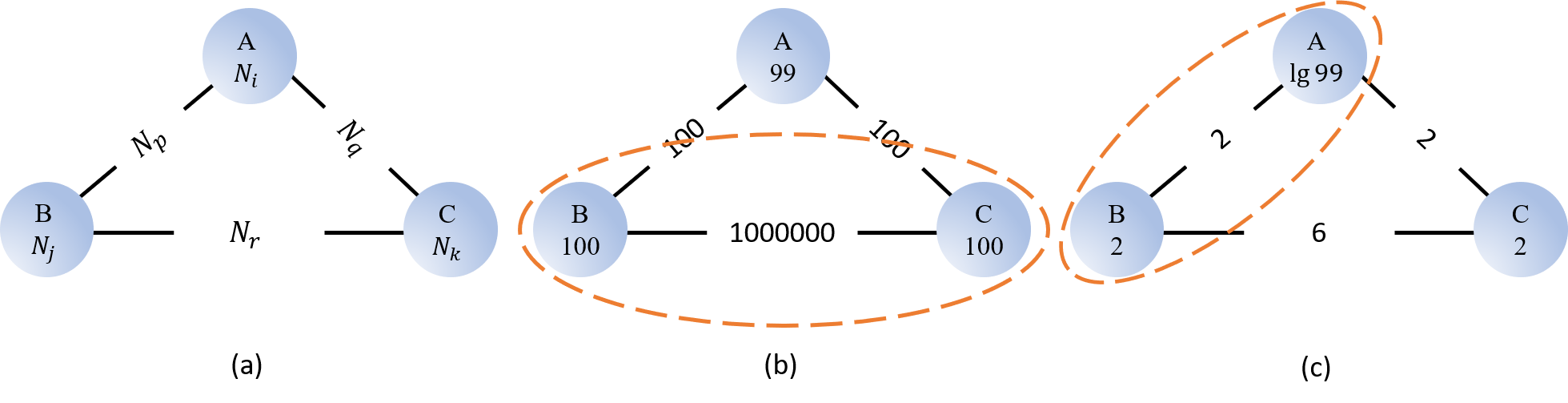}
    \caption{The difference between OMS and CMS for the same tensor network. (a) is the tensor network of equation \ref{dif_bet}. For (b) we adopt a multiplicational representation and construct an OMS problem. If we contract $A$ with $B$ at first and then $(AB)$ with $C$ (or symmetrically $A$ and $C$ first and then $B$), the total operation number is $99\times100\times100\times100\times1000000+99\times100\times100\times100\times100\times1000000 = 1.98\times 10^{14}$. Meanwhile, if we contract $B$ with $C$ and then $(BC)$ with $A$, the total operation number is $100\times100\times1000000\times100\times100 + 99\times100\times100\times100\times100 = 1.00\times 10^{14}$. As a result, the sequence $(BC)A$ is optimal in the OMS problem at part $(b)$. Now we adopt the additional representation to the same tensor network, by taking logarithm (with base 10) on the ranges of indices, and then we can construct a CMS problem as is shown in (c). The optimal sequence(s) on (c) is, however, different from that of (b). Consider the sequence $(BC)A$, and the $P_T$ number is $\max\{14, 8+lg99\}=14$. But for the sequence $(AB)C$ (or $(AC)B$), its $P_T$ number should be $\max\{12+lg99, 12+lg99\} = 12+lg99$, which is less than that of $(BC)A$. These results show that an OMS problem and a corresponding CMS problem may have totally different solutions.}
    \label{dif}
\end{figure}

The similarity of CMS and OMS are obvious: CMS minimizes the largest term while OMS minimizes the total, and the minimal operation number will be the same as the minimal time complexity under big O, given that the exponential base of additional representation is sufficiently large. However, we may also see the difference between CMS and OMS problem settings from Figure \ref{dif}. In this figure, the tensor network is actually the following equation:
\begin{equation}
\label{dif_bet}
(ABC)_{ijk} = \sum_{p=1}^{N_p}\sum_{q=1}^{N_q}\sum_{r=1}^{N_r}A_{ipq}B_{jpr}C_{kqr}
\end{equation}
. Suppose $N_i = 99$ be the range of index $i$, and the same for $N_j=100, N_k=100, N_p=100, N_q=100, N_r=1000000$. If we show the equation in an multiplicational representation, then it turns out to be  Figure \ref{dif} part (b), and if in a  additional representation then the part (c). Therefore, we may treat (b) as an OMS problem, and an optimal sequence should be $(BC)A$. Also, we may treat (c) as an OMS problem, and an optimal sequence should be either $(AB)C$ or $(AC)B$, but not the same as (b). This shows the difference of OMS and CMS settings, even for the same tensor network. 

From this example, we may see that the OMS problem, which optimizes the total operation number, is more accurate than its CMS version while concerning the optimization of computational costs. However, we also feel intuitively that the notation and calculation of CMS problem is much easier. There are important reasons that support us focusing more on CMS problem, and we will talk specifically in a following Section \ref{rel_CMS_OMS}.

\section{Proof of Module Degeneration}
{In this section we try to degenerate CMS problems to CMS-0 problems, which is a sub-problem of CMS. %We firstly prove that any sequence is optimal of a 3-vertex CMS-0 problem, and then propose the theorem that any vertex can be the last one to contract in an arbitrary CMS-0 problem. We also prove that for any step of contraction, we can contract the other vertices together in advance of this step. With those theorem, we can reduce CMS to CMS-0, by adding a new vertex and transforming the weights of vertices to those of edges connecting to this new vertex. 
With those theorem, we can reduce CMS to CMS-0 by adding a new vertex and transforming the weights of vertices to those of edges connecting to this new vertex.}

\textbf{Theorem 1}: {For a 3-vertex CMS-0 problem, any contraction sequence is a qualified solution.}

According to Theorem 1, if there are only three vertices in a CMS-0 graph, we can contract them directly without considering the sequence. In the following part, without ambiguity, we may contract 3 vertices at a time in a CMS-0 problem.

\textbf{Theorem 2}: {For any contraction sequence in CMS-0 problem, there exists one step other than the final step, whose $P_T$ is the largest among all the steps.}

Therefore, without losing generality, we do not necessarily consider the final step in CMS-0 problem. In other words, we only need to consider the $P_T$ of contractions that have at least one vertex not involved in the network. 
%This reveals that for every contraction we have to consider, there exist a group of vertices that are not involved. As a result, we can use a triad of 3 groups of vertices to represent an arbitrary contraction. For example, suppose $G_A$ and $G_B$ are the two groups of vertices that are going to be contracted in the following first step of a certain sequence. We also%%%%%%%%%%%%%%%%%%%%%%%%%%%%%%%$To be determined$%%%%%%%%%%

\textbf{Theorem 3}: {For a CMS-0 problem and any sequence to contract this network, and for any specific vertex in this network, there exist another sequence of contractions such that: (1) the two sequences are equal to each other in $P_T$; (2) the specific vertex can be the last one to be contracted in the newly-existing sequence.}

\textbf{Theorem 4}: {In a CMS-0 problem, for any sequence $Q_1$ of contractions and any step of contracting $V_1$ and $V_2$, there exists another sequence $Q_2$ that satisfies: (1) $Q_1$ and $Q_2$ are equal in $P_T$; (2) the steps of generating $V_1$ and $V_2$ in $Q_1$ are the same as that in $Q_2$; (3) in the sequence $Q_2$, for the vertices not in $V_1$ and $V_2$, we firstly contract them into a single ``vertex'' $V_3$, and then contract $V_1, V_2$ and $V_3$.}

\begin{figure}
    \centering
    \includegraphics[width = 0.9\textwidth]{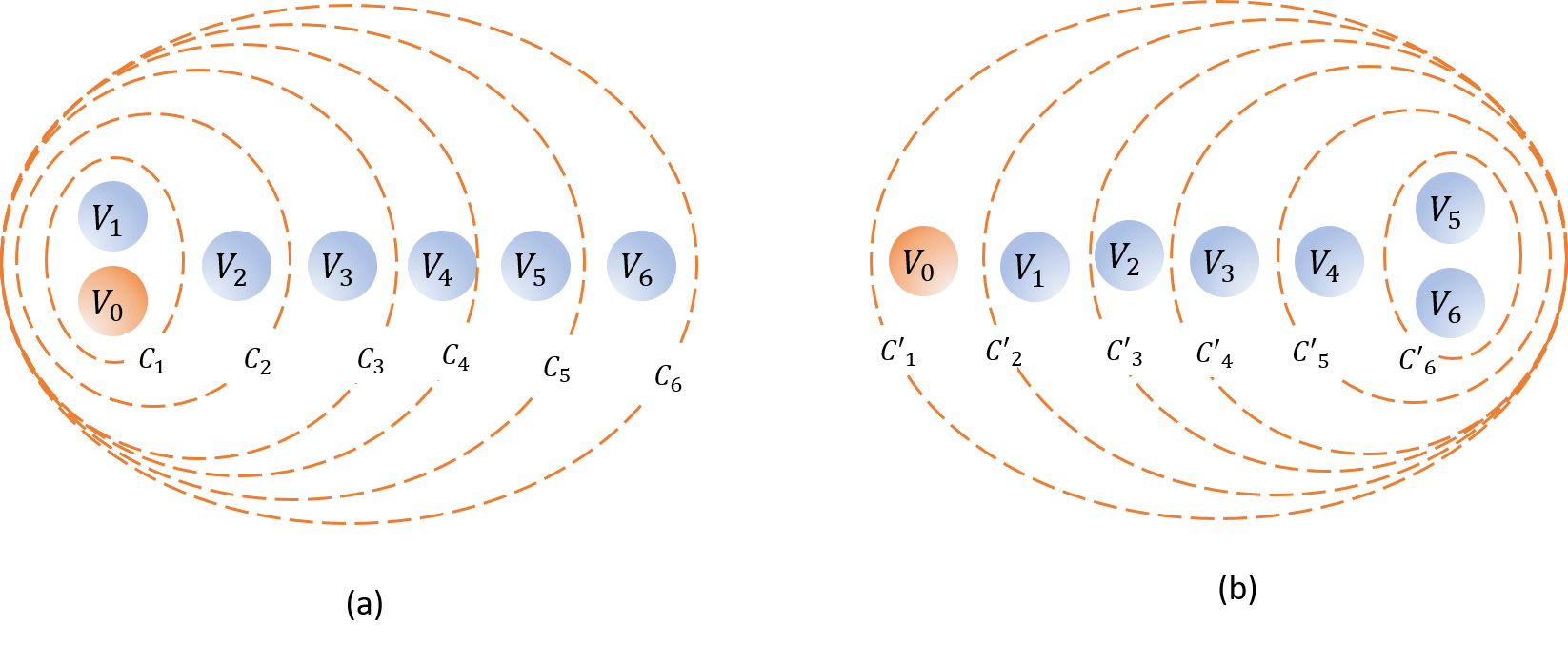}
    \caption{This is an illustration of Theorem 3 and Theorem 4. You may check and see that $P_T(C_i) = P_T(C'_i), \forall i =1,2,3,4,5,6$, corresponding to Theorem 3 that every vertex can be contracted in the last step. Also, consider the contraction of $V_0$ and $V_1$ in (a). If we firstly contract $V_2$ to $V_6$ into a whole vertex group and then contract it with $V_0$ and $V_1$, as in (b), the $P_T$ will not change.}
    \label{T34}
\end{figure}
Therefore, for any step of contraction, we can treat all vertices not involved as a whole group. You may also have an intuitive understand of Theorem 3 and Theorem 4 from Figure \ref{T34}. This comes with the following corollary:

\textbf{Corollary 1 }: {For any step of contraction in a CMS-0 problem, we can represent it with a combination of 3 groups of vertices $(P, Q, R)$, where $P$ and $Q$ are the 2 groups to be contracted, and $R$ is the group of other vertices.}

\textbf{Corollary 2}: {According to Theorem 1, under the condition given by Corollary 1, there are contractions $(P,Q,R)=(Q,R,P)=(Q,P,R)$.}

In the following parts, we will use the representation proposed in Corollary 1. Also, we will use the equations proposed in Corollary 2 to make arrangements on sequences. 
With these corollaries, we propose and prove the following theorems:

\textbf{Theorem 5}: {CMS can be reduced to CMS-0 within polynomial time.}

\begin{figure}
    \centering
    \includegraphics[width = 0.9\textwidth]{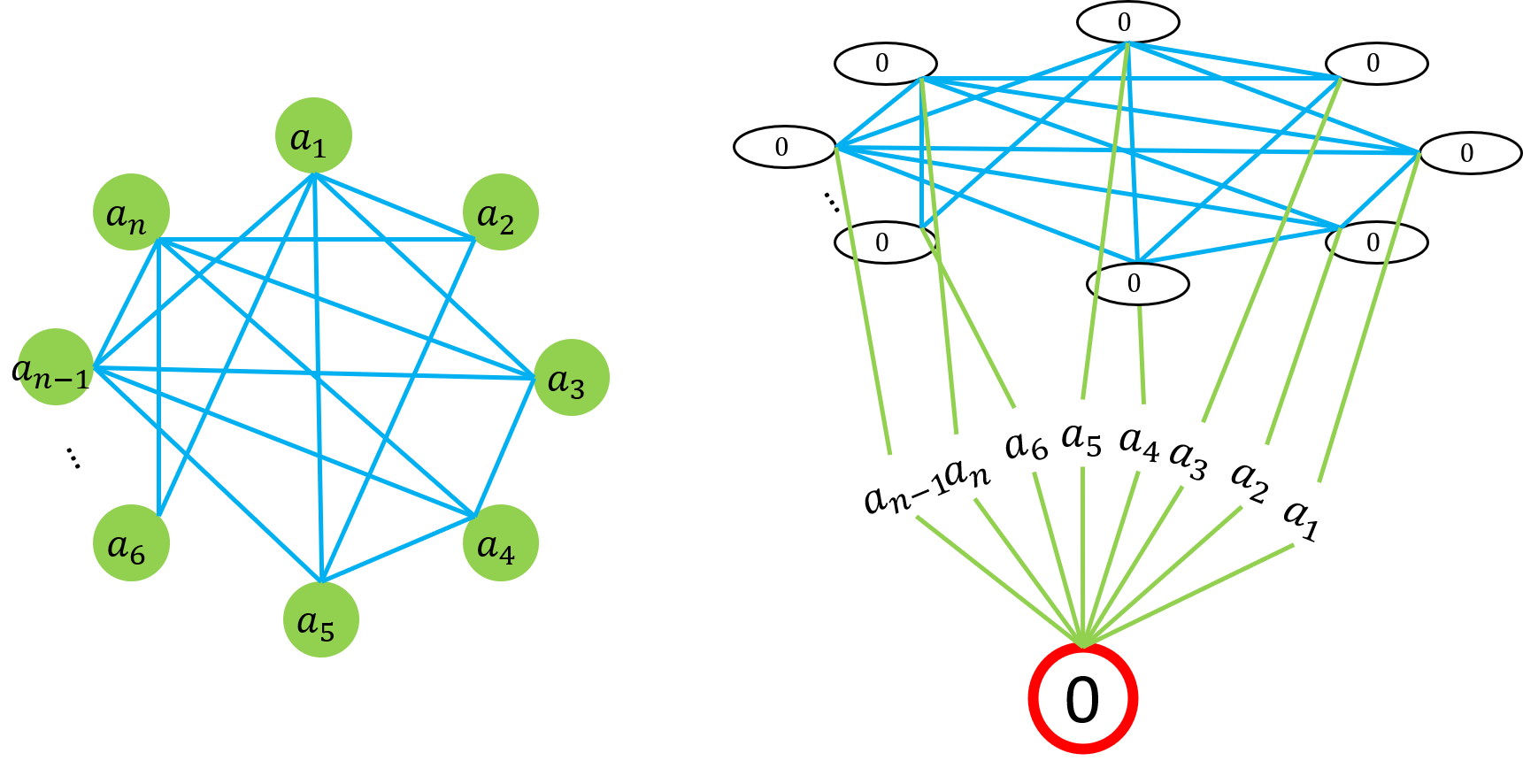}
    \caption{Reduction of CMS to CMS-0. Since CMS-0 is a subproblem of CMS, we only need to consider the transformation from CMS to CMS-0. This kind of transformation can be conduct by the following steps: (1) add a new vertex $V_0$ (the red vertex on the figure); (2) for any vertex $V_i , i=1,2,\ldots, n$, we connect $V_i$ with $V_0$ by an edge $e_{0,i}$; (3) we set the weight of each $e_{0, i}$ as the weight of $V_i$ (which is $a_i$ on the figure); (4) we set the weights of all vertices as 0.}
    \label{Tr}
\end{figure}

Detailed proof of Theorem 5 can be found in our appendix, and the general idea of this reductions is shown in Figure \ref{Tr}. For a CMS problem, we can transform it to an equivalent CMS-0 problem by adding a vertex and carefully setting the weights of edges connecting this vertex. Therefore, without losing generality, we can determine the solution of any CMS problem by determining its CMS-0 counterpart. Furthermore, if CMS-0 is proved to be a polynomial problem, then CMS is also a polynomial one. Also, since it is apparent that CMS-0 is a subproblem of CMS, we can state that CMS-0 can be reduced to CMS in polynomial time. That is to say, if CMS-0 is proved to be NP-hard, then CMS must be NP-hard as well. In a word, CMS and CMS-0 are equivalent to each other on the scale of polynomial and NP-Hardness.

\section{Why CMS is Easier than OMS}
\label{rel_CMS_OMS}
In this section we mainly prove 2 theorem on the relationship of CMS and OMS problem. The first reveals the relationship of their hardness, and the second indicates the relationship of their simpleness.

\textbf{Theorem 6}: {The condition that CMS is NP-hard is sufficient to that OMS is NP-hard.}

\textbf{Theorem 7}: For \textit{tensor tree networks}, the CMS problem is polynomial while the OMS is NP-hard.

With these two theorem above, we know that OMS is harder than CMS not only in general but also in tree cases. Also, with Theorem 7 individually, we can assert that OMS is NP-hard. %Since we are going to prove CMS-0 a NP-hard problem, we will know that CMS and OMS are both NP-hard.
\section{Proof of NP-Hardness}
{In this section we prove that CMS-0 is NP-hard, by reducing another NP-hard, the ``Exact Partition'' problem, to this CMS-0 problem. We will firstly construct a tensor network and figure out the optimal sequence in Theorem 8. After that, we point out that the exact partition problem is NP-hard in Theorem 9. Finally, in Theorem 10 we slightly change the tensor network in Theorem 8 such that it has the same properties of an optimal sequence, and reduce an exact partition problem to this CMS-0 problem.}

\textbf{Theorem 8}: {In CMS-0 problem settings, consider a tensor network with $(2n+1)$ vertices, where $n$ is large enough. Suppose the network is a complete graph with each edge weighs a same constant positive real number (or 1 without losing generality). Then a situation $(A,B,C)$ with $|A|=1, |B|=|C|=n$ must be in any optimal sequence.}

\textbf{Corollary 3}: {For the problem proposed in Theorem 8, there exists a sequence of contractions that has one and only one step of contraction whose $P_T$ value is  $(n^2+2n)$ and whose structure is $(A,B,C)$ with $|A|=1, |B|=|C|=n$. In fact, we can design the sequence $Q_0$ as follows: firstly contracting 2 vertices into a basic vertex, and for each time contract this basic vertex with another single vertex until no vertices left.} 
%%%

\textbf{Definition (Exact-Partition)}: Suppose there be $2n$ integers $A=\{a_1, a_2, \dots, a_{2n}\}$, and their sum equals $S$. We are asked to decide whether there exists $A' \subset A$, such that $|A'| = n$ and $\sum_{a_i \in A'}a_i = \frac{S}{2}$.

\textbf{Theorem 9}: {Exact-Partition problem is NP-complete.}

\textbf{Theorem 10}: {Exact-Partition problem can be reduced to CMS-0 problem.}%%%%%%%To Be Determined$$$$$$

Since CMS-0 is a subproblem of CMS, we know that CMS is NP-hard. Therefore, even though CMS is easier than OMS in the scale of reductions and on at least some specific structures of tensor networks, OMS and CMS are still both NP-hard.

\section{Conclusion and Open Problems}
{In this paper we have discussed the problem of optimally contracting tensor networks. We firstly conclude 2 different versions of this problem: OMS and CMS. Then we equivalently transform the CMS problem to its subproblem, CMS-0. Since OMS problem was proved NP-hard, we prove that CMS is easier than OMS both in general and in specific (tree) cases. However, we prove that the CMS problem is still NP-hard by reducing another NP-hard problem to CMS-0. 

In general, our work has settled a framework of tensor network contraction problems, by clearly determining and comparing the hardness of OMS and CMS problems both in general and in specific cases. Based on this work, not only the hardness would be released of designing an algorithm for a tensor network contraction problem, by transforming it from OMS to a corresponding and easier CMS problem, but also the operations would be simplified of contracting a CMS tensor network, by equivalently transforming it to a CMS-0 problem. Even though OMS is as NP-hard as CMS (and as CMS-0) in general, CMS is still easier in some specific cases, and CMS-0 is also a simplification of CMS problems.

There are, however, some open problems remaining. First of all, the proof in \cite{chi1997optimizing} is based on pairwise contractions. Even though it is proved in \cite{xu2019towards} that pairwise contraction can lead to an optimal solution of CMS problem, we can actually observe from the proof of Theorem 7 and find that pairwise contraction may not be an optimal choice under an OMS problem setting.  Secondly, for CMS (or equivalently, CMS-0) problems, we only know that they are NP-hard in some extremely hard cases, and polynomial in some extremely easy cases. We are curious about the boundary between easiness and hardness among CMS problems. Last but not least, our work has not propose conclusions on space complexities of tensor network contractions. The work  \cite{markov2008simulating} tries to determine a sequence with optimal space complexity, and they have reduced this problem to that of finding the treewidth of its \textit{line graph}. The problem of treewidth is in general NP-hard \cite{arnborg1987complexity}, but it is unsure to be true of any line graph. Even though the algorithm proposed in \cite{xu2019towards} guarantees a sequence optimal on both time and space complexity, we still cannot prove that the two problems are equivalent or approaching on the scale of hardness. Therefore, it is worth studying whether or not an optimal-space-complexity sequence on a tensor network is NP-hard .
\bigskip
}
\section*{Acknowledgement}
{
    We gratefully thank Mr. Pengfei Yu from Tsinghua University for the inspiration and contribution to the proof of Theorem 8!
}
\bibliography{ref}
\newpage
\section*{Appendex}
\begin{appendix}
\textbf{I Proof of Theorem 1}:

\textit{Proof}: {Actually, for different sequences of contracting a 3-vertex CMS-0 tensor network, the time powers for the first step are the same, which equal the sum of weights of the three edges.}${\blacksquare }$

\textbf{II Proof of Theorem 2}:

\textit{Proof}: {Consider the condition before the last but one step, and there should be three vertices. According to Theorem 1, we know that the time power of the first step equals the sum of weights of the three edges. Also, the time power of the last step equals the sum of weights of two edges. Therefore, the time power of the last step will not exceed that of the last but one step.}${\blacksquare }$

\textbf{III Proof of Theorem 3}: %{For a CMS-0 problem and any sequence to contract this network, and for any specific vertex in this network, there exist another sequence of contractions such that: (1) the $P_T$s of two sequences are equal to each other; (2) the specific vertex can be the last one to be contracted in the newly-existing sequence.}

\textit{Proof}: {Given a tensor network $G(V,E,W_e)$ and a specific vertex $V_0$, and for an arbitrary sequence $Q_1$ of contractions, we construct another sequence $Q_2$ that satisfies the theorem.}

{Consider the condition before the last but two step of contraction, where there are three ``vertices'' remaining. Suppose these 3 ``vertices'' to be $V_1, V_2, V_3$, and suppose that $V_0 \in V_1$. Define $x\triangleq |V_1|$.} 

{If $x = 1$, then we can construct $Q_2$ by firstly contracting the other two vertices, and then contract $V_0$ with the combined vertex. According to Theorem 1 and Theorem 2, we know that this adjustment would not change the $P_T$ value. Therefore, $P_T(Q_2)=P_T(Q_1)$.}

{Otherwise, $x > 1$. We now construct another sequence $Q_1^{'}$ such that: for the condition before the last but one step in $Q_1^{'}$, $V_0$ is in one of the three ``vertices'' $V_1^{'}$, and $|V_1^{'}| < x$. In fact, suppose that $V_1$ is the contraction of $V_{10}$ and $V_{11}$. Since the contractions within $V_2$ or $V_3$ have nothing to do with those within $V_1$, we can assume that the last but three step is to contract $V_{10}$ and $V_{11}$. Now we construct $Q_1^{'}$ as follows: we swap the contraction of $V_{10}$ and $V_{11}$ with the contraction of $V_2$ and $V_3$. Note that these two contractions are independent. Therefore, the swap will not change the $P_T$ value, which means that $P_T(Q_1)=P_T(Q_1^{'})$. Also, either $V_{10}$ or $V_{11}$ contains $V_0$. Since $|V_{10}|+|V_{11}|=x$, and $|V_{10}|\geq1, |V_{11}|\geq1$, we have that $|V_{10}|< x, |V_{11}|< x$. Thus, we find out a qualified $Q_1^{'}$.}

{By induction on the $x$, and we can iteratively reduce $x$ by at least 1 in each time until $x=1$, which leads to the first case. Thus the theorem holds.}${\blacksquare }$

\textbf{IV Proof of Theorem 4}:%{In a CMS-0 problem, for any sequence $Q_1$ of contractions and any step of contracting $V_1$ and $V_2$, there exists another sequence $Q_2$ that satisfies: (1) $Q_1$ and $Q_2$ are equal in $P_T$; (2) the steps of generating $V_1$ and $V_2$ in $Q_1$ are the same as that in $Q_2$; (3) in the sequence $Q_2$, for the vertices not in $V_1$ and $V_2$, we firstly contract them into a single ``vertex'' $V_3$, and then contract $V_1, V_2$ and $V_3$.}

\textit{Proof}: {In fact, if we consider the result of contraction of $V_1$ and $V_2$, which is a ``vertex'' $V_{12}$ after contraction, we can simply apply Theorem 3 so that $V_{12}$ can be contracted in the final step. Consider every steps after the contraction of $V_1$ and $V_2$ and before the final step, and those steps contract the vertices other than those in the 2 groups into one vertex (group), which is exactly the $V_3$. Those steps do not involve $V_{12}$, and also generating $V_1$ and $V_2$ does not involve $V_3$ As a result, we can firstly generate $V_3$ and then contract $V_1, V_2$ and $V_3$. Thus, we construct a $Q_2$ satisfying all the three conditions above. These adjustments will not change the time power of the sequence.}${\blacksquare }$

\textbf{V Proof of Theorem 5}: {CMS can be reduced to CMS-0 within polynomial time.}

\textit{Proof}: {For a tensor network $G(V,E,W_v,W_e)$ in the CMS problem, we can construct another tensor network $G^{'}(V^{'},E^{'},W^{'}_e)$ in polynomial time, such that for any contraction sequence $Q_1$ on $G$, we can find out a contraction sequence $Q^{'}_1$ on $G'$ such that $P_T(Q_1)$ equals $P_T(Q^{'}_1)$, and vice versa. In fact, we can construct $G^{'}$ from $G$ by simply add a vertex $V_0$, and for any vertex $v$, we assign $W_{v-V_0}=W_{v}$, then we remove $W_{v}, \forall v\in V(G)$. According to Theorem 3, we can suppose $V_0$ to be contracted only in the last step without changing the $P_T$ value. Therefore, for the contraction steps before the last step on $G^{'}$, they have the same time powers as the corresponding contraction steps on $G$. Therefore, the CMS-0 problem is equivalent to the original CMS problem.} ${\blacksquare }$

\textbf{VI Proof of Theorem 6}: %{The condition that CMS is NP-hard is sufficient to that OMS is NP-hard.}

\textit{Proof}: {%We prove this theorem by contradiction. First of all, if the dimension of each order of tensors is sufficiently large, then the solution of an OMS problem must be a solution of the corresponding CMS problem (otherwise any small increase on the time power will lead to extreme increase on operation number). If we suppose that OMS is not NP-hard, then there exist a polynomial algorithm that can solve OMS problem, and thus
We prove the following proposition: for a CMS problem, we can find a corresponding OMS problem in polynomial time such that the solution to this OMS problem is also one solution to the CMS problem. Apparently, given this proposition, and given that the CMS problem is NP-hard, the corresponding OMS problem is NP-hard, and thus the OMS problem is NP-hard as well.

Given a tensor network $G(V,E,W_V,W_E)$ for a CMS problem, $|V|\geq 3$, we now build up a corresponding $G^{'}(V',E',W'_{V'},W'_{E'})$ problem according to the follows:

\begin{enumerate}[i]
	\item $V'(G')=V(G)$.
	\item $E'(G')=E(G)$.
	\item For any vertex $v\in V'$ (also $v\in V$), we set the weight $W'_{v} = N^{W_v}$, where $N \in \mathbb{N}^(+)$.
	\item For any edge $e\in E'$ (also $e\in E$), we set the weight $E'_{e} = N^{W_e}$, where $N$ is the same as that in (iii).
\end{enumerate}

According to this construction, for any step of contraction $C_1$ on $G$ and, correspondingly, on $G'$, we know that $OPN(C_1)$ in $G'$ equals $N^{P_T(C_1)}$ in $G$. Therefore, suppose $Q_1$ is an optimal solution to the OMS problem on $G'$, and then we know:

$$\frac{OPN(Q_1)}{n}\leq N^{P_T(Q_1)}\leq OPN(Q_1)$$
, where $n=|V|=|V'|$. Now, let us determine the value of $N$:

Define $\Delta = min_{{W_1}, {W_2}\subseteq W_V\cup W_E, S(W_1)\neq S(W_2)} (S(w_1) - S(w_2))$, where $S(A)$ denotes the sum of all elements in set $A$, and then assign $N = max\{n^{\frac{2}{\Delta}}, 2 \}$. If $Q_1$ is not a solution to the corresponding CMS problem, then there at least exists another sequence $Q_2$ such that $P_T(Q_2)<P_T(Q_1)$. According to the definition of $\Delta$, we know that $P_T(Q_2)\leq P_T(Q_1)-\Delta$Therefore, we have:
$$N^{P_T(Q_2)}\leq \frac{N^{P_T(Q_1)}}{N^{\Delta}}\leq \frac{OPN(Q_1)}{N^\Delta}$$
. Note that $N^\Delta=(max\{n^\frac{2}{\Delta}, 2 \})^\Delta\geq (n^\frac{2}{\Delta})^{\Delta}=n^2$, we have:
$$N^{P_T(Q_2)}\leq\frac{OPN(Q_1)}{n^2}$$
. However, we also have that $OPN(Q_2)\leq n\times N^{P_T(Q_2)}$. Therefore, we have $OPN(Q_2)\leq n\times \frac{OPN(Q_1)}{n^2}<OPN(Q_1)$. This is contradict to the definition of $Q_1$, whose operation number is the minimal. Thus, the proposition holds, and so does the theorem.

}${\blacksquare }$

\textbf{VII Proof of Theorem 7}: % For \textit{tensor tree networks}, the CMS problem is polynomial while the OMS is NP-hard.

\textit{Proof}:
{
On one hand, the CMS problem on tree is proved to be polynomial by the work \cite{xu2019towards}. Now, on the other hand, we propose an OMS problem on a tensor tree network that is NP-hard.

Consider the following case: there are $(n+1)$ vertices in the tensor network, and we denote them as $V_0, V_1, \ldots, V_n$. Suppose $V_0$ is connected to all $V_1, V_2, \ldots, V_n$, and $V_i$ is only connected to $V_0$, $i=1,2,3,\ldots,n$. Denote the edge connecting $V_0$ and $V_i$ as $e_i$, and denote the weight of $e_i$ as $b_i$, $i=1,2,3,4,\ldots,n$. Suppose the weight of $V_0$ to be $a$, and the weight of $V_i$ to be $1$, $i=1,2,...,n$.

Since we know that in OMS problems the weight of edges are all integers, we have $b_i \geq 1$. Without losing generality, we can suppose that $n \geq 2$. Now we set $a=2n\prod_{i=1}^{n}b_i$. In this case, for each step of contraction, if this step involves $V_0$, then the operation number will definitely exceed $a$. Consider 2 kinds of sequences of contractions: $Q_1$ has only one step involving $V_0$, which firstly does contractions among $V_i, i=1,2,\ldots, n$, and finally contract the combination of $V_i$ with $V_0$; $Q_2$ has at least 2 steps involving $V_0$. Therefore, it shows that

$$OPN(Q_1)\leq a\times {\prod_{i=1}^{n}b_i} + n\times{\prod_{i=1}^{n}b_i}$$
, and that
$$OPN(Q_2)\geq a + a \times \prod_{i=1}^n b_i$$
. Since $a=2n\prod_{i=1}^{n}b_i>n\prod_{i=1}^{n}b_i$, we have that $OPN(Q_1)<OPN(Q_2)$. Therefore, the optimal sequence must be one of $Q_1$s. In other words, we only need to figure out the optimal sequence of contracting $V_1, V_2, \ldots, V_n$. Since the last step of contracting $V_1, V_2, \ldots, V_n$, which is also the last but one step of contracting the whole tensor network, has the operation number of $\prod_{i=1}^{n}b_i$, we only need to optimize the operation number in the previous steps. 

Let $b_i = Mb'_i$ for all $i \in [n]$, where $M = 2n\prod_{i = 1}^n b'_i$, and let $S=\{b_1, b_2, \ldots, b_n\}$, $T=\{1,2,...,n\}$. Let $n$ be an even integer. Without loss of generality, assume we contract $V_t$ and $V_{t'}$ and generate $V_T$, where $t \subseteq T$ and $t' = T \setminus t$, and $|t| \geq |t'|$. If $|t| = |t'| = n/2$, then in each step before generating $V_T$, the operation number is at most $M^{n/2}\prod_{i = 1}^n b'_i$, thus the total operation number is at most $nM^{n/2}\prod_{i = 1}^n b'_i$. But if $|t| > n/2$, to form $V_t$, we need at least $M^{|t|} \geq M^{n/2 + 1} = 2nM^{n/2}\prod_{i=1}^n b'_i$ operations, which is strictly greater then the previous case. Thus in the optimal contraction sequence, $|t| = |t'| = n/2$. Now consider the contraction steps before generating $V_t$ and $V'_t$, in each step, we need at most $M^{n/2-1}\prod_{i=1}^n b'_i$ operations, and to form $|V_t|$ and $V_{t'}$, we need $(\prod_{i \in t}b'_i + \prod_{j \in t'}b'_j)M^{n/2}$ operations, thus the total number of operations we need before contracting $V_t$ and $V_{t'}$ is at most $(\prod_{i \in t}b'_i + \prod_{j \in t'}b'_j)M^{n/2} + nM^{n/2-1}\prod_{i=1}^n b_i \leq (\prod_{i \in t}b'_i + \prod_{j \in t'}b'_j + 1/2)M^{n/2}$. As $b_i$s are integers, so we just need to minimize $\prod_{i \in t}b'_i + \prod_{j \in t'}b'_j$ in this step, and otherwise the total operation number cannot be optimal. And also we have $\prod_{i \in t}b'_i + \prod_{j \in t'}b'_j \geq 2\sqrt{\prod_{i=1}^n b'_i}$, and the equal sign holds if and only if $\prod_{i \in t}b'_i = \prod_{j \in t'}b'_j$, thus we can reduce SPPF problem to OMS problem according to the previous argument. 

\textbf{Strictly Partition in Product Form(SPPF)}: given a set of $n$ positive integers $\{b_1, b_2, \dots, b_n\}$, where $n$ is even, decide whether there exists $T \subsetneq S$, such that $\prod_{b \in T}b = \prod_{b \in S \setminus T}b$, and $|T| = n/2$. 

In the reduction, we have only polynomial times arithmetic operations on the input of the original OMS problem, the reduction can be done in polynomial time. And later I'll show that SPPF is NP-Hard, thus OMS is also NP-Hard. As OMS is in HP, so OMS is NP-Complete. Consider the following two problems. 

\textbf{Partition in Product Form(PPF)}: given a set $S$ of $n$ positive integers $\{b_1, b_2, \dots, b_n\}$, decide whether there exists $T \subseteq S$, such that $\prod_{b \in T}b = \prod_{b \in S \setminus T}b$. 

\textbf{Subset Product(SP)}: given a set $S$ of $n$ positive integers $\{b_1, b_2, \dots, b_n\}$ and $K > 1$, decide whether there exists a subset $T$ of $S$, such that $\prod_{b_i \in T}b_i = K$.  

Lemma 1: SP is NP-Complete. \cite{book:340980}

Lemma 2: SP can be reduced to PPF in polynomial time. 

Proof: Let $N = \prod_{i = 1}^n b_i$. Consider the PPF problem on the set $S' = S \cup \{N, K^2\}$, if there exists $T \subseteq S$ such that $\prod_{b \in T}b = K$, then let $T' = T \cup \{N\}$, then we have $\prod_{b \in T'}b = NK = \prod_{b \in S'\setminus T'}b$. And if there exists $T' \subseteq S'$ such that 
$\prod_{b \in T'}b = NK = \prod_{b \in S'\setminus T'}b$, as all the elements are at least 1, $N$ and $K^2$ can not be in the same side, thus only one of $N$ and $K^2$ are in $T'$. Without loss of generality, assume $N \in T'$, then let $T = T' \setminus \{N\}$, as $K^2 \notin T$, we know $T \subseteq S$, and $\prod_{b \in T}b = \prod{b \in T'}b / N = K$. Thus the SP instance is satisfiable if and only if the corresponding PPF instance is satisfiable. And we  just need polynomial times arithmetic operations in the reduction, so this reduction can be done in polynomial time. 

Lemma 3: PPF can be reduced to SPPF in polynomial time. 

Proof: For a PPF instance on $S = \{b_1, b_2, \dots, b_n\}$, consider the SPPF problem on $S' = \{b_1, b_2, \dots, b_n, b'_1, b'_2, \dots, b'_n\}$, where $b'_i = 1$ for all $i \in [n]$. If there exists $T' \subseteq S'$ such that $|T'| = n$ and $\prod_{b \in T'}b = \prod_{b \in S' \setminus T'}b$, then we have $\prod_{b \in T' \cap S}b = \prod_{b \in (S' \setminus T') \cap S}b$, as all the added elements are 1, thus there exists a PPF on $S$. If $S$ has a PPF $T \subseteq S$, then we can pad $n - |T|$ ones into $T$ and form $T'$, which will become a SPPF on $S'$. Thus PPF can be reduced to $SPPF$ in polynomial time. 

(\textit{Remark:} here the OMS problem is based on pairwise contractions. In fact, if we contract $V_0, V_1, \ldots, V_n$ all together simultaneously, then the operation number is even smaller than that of an optimal sequence of pairwise contractions.)
%Similar to the proof of Theorem 6, we can select $N$ to be sufficiently large. In other words%In other words, we can assume that the last but one step of contracting the set $V_T$ has the largest operation number among all steps before the last one. 

%Under this assumption, a necessary condition of minimizing the total operation number is to minimize that of 

}${\blacksquare }$

\textbf{VIII Proof of Theorem 8}: %{In CMS-0 problem settings, consider a tensor network with $(2n-1)$ vertices, where $n$ is large enough. Suppose the network is a complete graph with each edge weighs a same constant positive real number (or 1 without losing generality). Then a situation $(A,B,C)$ with $|A|=1, |B|=|C|=n$ must be in any optimal sequence.}

\textit{Proof}:{
	Suppose $G(V,E, W_E)$ be the tensor network of a CMS-0 problem, where $W_e = 1, \forall e\in E$. Suppose $Q_1$ is an optimal sequence of contractions of $G$, and denote a step $S_1$ to be that of the largest $P_T$ value in the sequence $Q_1$. Suppose $S_1 = (A, B, C)$, where $A, B, C\subset V, A\cap B = B\cap C=C\cap A=\emptyset, A \cup B \cup C = V$. Suppose $a = |A|, b=|B|, c=|C|, a\leq b\leq c, a+b+c=2n+1$. Without losing generality, we also let $S_1$ have \textbf{a smallest $a$} among all largest-$P_T$ step(s) in $Q_1$. Therefore, we consider the following cases:

\begin{enumerate}[i]
	\item If $a=1$, then $b\leq n, c\geq n$. Suppose the group $C$ is contracted from 2 sup-groups $C_1$ and $C_2$. Therefore, consider the step of contracting $C_1$ and $C_2$, denote this step as $S_2$, and we will have:
	$$P_T(S_2) =(a+b)(c_1+c_2)+c_1\cdot c_2= (a+b)\cdot c+c_1\cdot c_2$$
	, where $c_1=|C_1|, and c_2=|C_2|$. Since we have $c_1\geq1, c_2\geq1$, we know that $c_1\cdot c_2=(c_1-1)\cdot(c_2-1)+c_1+c_2-1\geq c_1+c_2-1 = c-1$. Therefore, if $b< c$, then $c\geq b+2$, and then
	\begin{equation}
	\begin{aligned}
	P_T(S_2)&=(a+b)c+c_1 c_2\\
	&\geq(a+b)c+c-1\\
	&>(a+b)c+c-2\\
	&\geq(a+b)c+b\\
	&=(a+b)c+ab\\
	&=P_T(S_1)
	\end{aligned}
	\end{equation}
, which contradicts with the assumption that $S_1$ is the step with a largest $P_T$ value. Therefore, $b=c=n$, and the theorem holds.
	
	\item If $a \geq 2$, suppose $A$ is contracted from the 2 groups of vertices $A_1$ and $A_2$, and denote $a_1 = |A_1|, a_2 = |A_2|$. Without losing generality, we suppose $a_1\leq a_2$. As a result, we have $a_1\leq a_2 < b\leq c$ In the following part, we are going to achieve a contradiction. 

	Consider the step before the contraction of $A_1$ and $A_2$, and we find that there are 4 groups of vertices $A_1$, $A_2$, $B$ and $C$ left. In $Q_1$ we firstly contract $A_1$ with $A_2$ and then contract $B$ and $C$. Now we construct another sequence $Q_2$ by keeping all steps in $Q_1$ before the last but three steps, then contract $A_2$ with $B$, then $A_1$ with $C$, and then the two groups together. Therefore, there are only 2 steps different between $Q_1$ and $Q_2$: $(A_1, A_2, (B+C)), (A, B, C)$ in $Q_1$, and $(A_1, C, (A_2+B)), ((A_1+C), A_2, B)$ in $Q_2$. Since $a_1\leq a_2 < b\leq c$, we have:
\begin{equation}
\begin{aligned}
P_T(A, B, C)&>P_T(A_1, C, (A_2+B))\\
P_T(A, B, C)&>P_T((A_1+C), A_2, B)\\
P_T(A, B, C)&>P_T(A_1, A_2, (B+C))
\end{aligned}
\end{equation}
. As a result, $P_T(Q_2)\leq P_T(Q_1)$, and $Q_2$ remove one of the largest-$P_T$ steps from $Q_1$ without generating new ones. Since $a\geq2$, all largest-$P_T$ steps in $Q_1$ can be decomposed and re-contracted as we have done above. Therefore, we can repeat doing this until all largest-$P_T$ steps are removed. We denote the current sequence as $Q_3$, and then we have $P_T(Q_3)<P_T(Q_1)$, which is contradict to the assumption that $Q_1$ is optimal. Thus $a\geq2$ leads to contradictions. 
\end{enumerate}

In conclusion, the theorem holds.
}${\blacksquare }$
%%%%%%%%%%%%%%%%%%%%%%%#$%@^$^@%&*&^%

\textbf{IX Proof of Theorem 9} %Suppose there be $2n$ integers, and their sum equals to $S$. We are asked to decide whether there exist $n$ of those numbers whose sum is $\frac{S}{2}$. Exact-Partition problem is NP-complete.

\textit{Proof}:
{
We reduce the Partition problem to this Exact-Partition problem.

Consider the partition problem on $A=\{a_1, a_2, \ldots, a_n\}$, we construct another set: $B=\{a_1+S, a_2+S, \ldots, a_n+S, S, S, \ldots, S\}, |B|=2n$. We will show that the partition problem on $A$ can be reduced to the Exact-Partition problem on $B$:

On one hand, if there exists a solution $A_1\subseteq A$ to the partition problem on $A$, then we have:
$$\sum_{a_i\in A_1} a_i = \frac{S}{2}$$
. Now we construct a solution $B_1$ to the exact partition problem on $T$:
$$B_1 = \{a_i+S|a_i\in A_1\}\cup \{S, S, \ldots, S\}, s.t. |B_1|=n$$
. We then have:

\begin{equation}
\begin{aligned}
\sum_{x\in B_1} x &= \sum_{a_i\in A_1} (a_i + S) + S(|B_1|-|A_1|)\\
&=\sum_{a_i\in A_1} a_i + S|A_1| + S(|B_1|-|A_1|)\\
&=\frac{S}{2} + S|B_1|\\
&=\frac{S}{2} + Sn\\
&=\frac{\sum_{x\in T} x}{2}
\end{aligned}
\end{equation}
. Thus $B_1$ is a solution to the exact-partition problem on $T$.

On the other hand, if there exist a solution $B_2$ to the exact-partition problem on $B$, then we construct a solution to the partition problem on $A$: $A_2\triangleq \{a_i|(a_i+S)\in B_2\}$. Similarly, we have $\sum_{a_i\in A_2} a_i = \sum_{x_i\in B_2}x - S|B_2| = \frac{S}{2}$. Therefore, $A_2$ is a partition on $A$.

Since all those operations above can be carried out within polynomial time, we have a polynomial reduction from the partition problem to the exact-partition problem. 
}${\blacksquare }$

\textbf{X Proof of Theorem 10} %Exact-Partition problem can be reduced to CMS-0 problem.

\textit{Proof}:

{Consider an exact-partition problem $A=\{a_1, a_2, \ldots, a_{2n-1}, a_{2n}\}$. Without losing generality, assume that $a_i >0, i = 1,2,\ldots,2n-1, 2n$. Denote $s\triangleq\sum_{i=1}^{2n}a_i$, and suppose $a_0 = \frac{s}{2}$. Now, let us define a CMS-0 tensor network $G(V,E,W)$, where:
$$V=\{v_0, v_1, v_2, \ldots, v_{2n-1}, v_{2n}\}$$
$$E=\{e_{i,j}|i,j\in\{0,1,2,\ldots, 2n \}, i\neq j\}$$

$$W=\{w_{i,j}|i,j\in\{0,1,2,\ldots, 2n \}, i \neq j\}$$
$$w_{i,j}=x-{a_{i}}{a_{j}}$$
, where $x$ is a sufficiently large positive number that we will determine later. Consider 2 groups of vertices $V_1=\{v_i|i\in N_1\}$ and $V_2=\{v_j|j\in N_2\}$, where $N_1 $ and $N_2$ are the index sets of $V_1$ and $V_2$, $N_1, N_2 \subset \{0,1,2,...,2n-1,2n\}, N_1\cap N_2 = \emptyset$. Then the sum of weights of edges between $V_1$ and $V_2$ is $|N_1|\cdot|N_2|-(\sum_{i\in N_1}a_i)(\sum_{j\in N_2}a_j)$.

Now, let's consider the CMS-0 problem on $G(V,E,W)$. Suppose $Q_1$ be an optimal sequence of contractions. Suppose a step in $Q_1$ be $S_1=(V_1, V_2, V_3)$, where $V_r={a_i|i\in N_r}, r=1,2,3$ and where $N_1, N_2, N_3\subset \{0,1,2,...,2n\}, N_1\cap N_2\cap N_3 = \emptyset, 1\leq|N_1|\leq|N_2|\leq|N_3|<2n$. Suppose $u=\sum_{i\in N_1}a_i, v=\sum_{j\in N_2}a_j, w=\sum_{k\in N_3}a_k$, and then we have:
$$u+v+w=a_0+\sum_{i=1}^{2n}a_i=\frac{s}{2}+s=\frac{3s}{2}$$
. Therefore, we have:
$$P_T(S_1)=x\cdot(|N_1|\cdot|N_2|+|N_2|\cdot|N_3|+|N_3|\cdot|N_1|)-(uv+vw+wu).$$

Consider another CMS-0 tensor network $G'(V',E',W')$, where $V'=V, E'=E, w_{i,j}=x, i,j\in\{0,1,2,...2n-1,2n\}, i\neq j$. Suppose $x$ be sufficiently large. If we apply $Q_1$ on $G'$, then there are 2 cases as follows:

\begin{enumerate}[i]
	\item $Q_1$ on $G'$ is also an optimal sequence, then we can pick up an $S_1$ such that $|N_1|=1, |N_2|=|N_3|=n$ according to Theorem 8. In this case, we have 
$$P_T(S_1)=x\cdot(1\cdot2n+n\cdot n-(uv+vw+wu))=x\cdot(n^2+2n)-(uv+vw+wu)$$
in the original tensor network $G$.
	\item $Q_1$ on $G'$ is not an optimal sequence, then we can find out an $S_2$ such that $P_T(S_2)\geq x(n^2+2n+1)$ on $G'$ according to the contrapositive of Theorem 8. Therefore, if we apply $S_2$ on the original tensor network $G$, and then we have:
$$P_T(S_2)\geq x(n^2+2n+1)-(u'v'+v'w'+w'u')$$
. As a result, we compare $P_T(S_1)$ with $P_T(S_2)$ and then we will have:
$$P_T(S_2)-P_T(S_1)\geq x -(u'v'+v'w'+w'u'-(uv+vw+wu))$$
. Since that $x$ can be sufficiently large, and that the combination of $u', v', w', u, v, w$ is finite, we can choose $x = (\sum_{a_i \in A}a_i + 1)^3 + 1 > (u'v'+v'w'+w'u'-(uv+vw+wu)) + 1, \forall u, v, w, u', v', w'$. Therefore, $P_T(S_2)>P_T(S_1)$, and thus $P_T(Q_1)>x\cdot(n^2+2n)-(uv+vw+wu)$ if $Q_1$ is not optimal on $G'$.
\end{enumerate}

To draw a conclusion, we have $P_T(S_1)\geq x\cdot(n^2+2n)-(uv+vw+wu)$, and the ``$=$'' holds only if $Q_1$ is also an optimal sequence on $G'$.  Note that:

$$uv+vw+wu\leq\frac{(u+v+w)^2}{3}=\frac{(\frac{3}{2}s)^2}{2}=\frac{3s^2}{4}$$
, and the ``$=$'' holds if and only if $u=v=w=\frac{s}{2}$. Since that $|N_1|=1, |N_2|=|N_3|=n$, and that $a_0=\frac{s}{2}, a_i > 0, \forall i\in\{0,1,2,\ldots, 2n\}$, we know that $N_1=\{0\}$ if $u=v=w=\frac{s}{2}$. That is to say, $\sum_{j\in N_2}a_j = \sum_{k\in N_3}a_k, N_2\cup N_3=\{1,2,...,2n\}$. Therefore, we have:

\begin{equation}
P_T(S_1)\geq x\cdot(n^2+2n)-\frac{3s^2}{4}
\end{equation}
, and the ``$=$'' holds if and only if (1) $Q_1$ is optimal on $G'$ and (2) $\sum_{j\in N_2}a_j = \sum_{k\in N_3}a_k, N_2\cup N_3=\{1,2,...,2n\}$. 

Therefore, if the CMS-0 problem on $G(V,E,W)$ has an optimal sequence $Q_1$ such that $P_T(Q_1)=x\cdot(n^2+2n)-\frac{3s^2}{4}$, then the 2 necessary conditions above must be satisfied. As a result, there must exist 2 sets of indices $N_2$ and $N_3$, $N_2, N_3\subset\{1,2,\ldots, 2n-1, 2n\}, |N_2|=|N_3|=n, N_2\cup N_3=\{1,2,...,2n\}$, such that $\sum_{j\in N_2}a_j = \sum_{k\in N_3}a_k$. Then $N_2$ and $N_3$ serves as a solution of the exact-partition problem on $A=\{a_1, a_2, \ldots, a_{2n-1}, a_{2n}\}$.

Also, if the exact-partition problem on $A$ has a solution $(A_1, A_2)$, then we construct a solution to the CMS-0 problem. Let $Q_1$ be the sequence proposed in Corollary 3, and $S_1$ be the only step that has the largest $P_T$ and correspondingly a structure of $(1,n,n)$. Since the sequence in Corollary 3 does not specify the vertices to be contracted, we can choose $N_1, N_2, N_3$ according to their specific size. If we specify $S_1$ by choosing the set $N_1=\{0\}, N_2=\{j|a_j\in A_1\}, N_3=\{k|a_k\in A_2\}$, then we know that $P_T(S_1)=x\cdot(n^2+2n)-\frac{3s^2}{4}$. Also, according to Corollary 3, for any $S_0\in Q_1, S_0\neq S_1$, we have $P_T(S_0)\leq x\cdot(n^2+2n-1)-(u_0v_0+v_0w_0+w_0u_0)$. Since $x$ is sufficiently large, we can assume that $P_T(S_0)<P_T(S_1)$. Therefore, $P_T(Q_1)=P_T(S_1)=x\cdot(n^2+2n)-\frac{3s^2}{4}$. Thus, $Q_1$ is a solution to the CMS-0 problem.

In a nutshell, we reduce the exact-partition problem to a CMS-0 problem.
}${\blacksquare }$. 

\end{appendix}
%\printbibliography%%%%%%%%%%%%%%
\end{document}